\documentclass[pdflatex,referee,sn-nature]{sn-jnl}% Math and Physical Sciences Numbered Reference Style 

\usepackage{graphicx}%
\usepackage{multirow}%
\usepackage{amsmath,amssymb,amsfonts}%
\usepackage{amsthm}%
\usepackage{mathrsfs}%
\usepackage[title]{appendix}%
\usepackage{xcolor}%
\usepackage{textcomp}%
\usepackage{manyfoot}%
\usepackage{booktabs}%
\usepackage{algorithm}%
\usepackage{algorithmicx}%
\usepackage{algpseudocode}%
\usepackage{listings}%
\usepackage{bm}%
\usepackage{caption}
\usepackage{ulem} %ADDED by CE
\theoremstyle{thmstyleone}%
\theoremstyle{thmstyletwo}%
\theoremstyle{thmstylethree}%
\begin{document}

\title[Coherent phonon control beyond amplitude saturation in a sliding ferroelectric]{Coherent phonon control beyond amplitude saturation in a sliding ferroelectric}


\author*[1,2]{\fnm{Jan Gerrit} \sur{Horstmann}}\email{gerrit.horstmann@uni-wuerzburg.de}

\author[3]{\fnm{Christoph} \sur{Emeis}}

\author[1]{\fnm{Andrin} \sur{Caviezel}}

\author[4]{\fnm{Quintin N.} \sur{Meier}}

\author[1]{\fnm{Nicolas} \sur{Wyler}}

\author[1]{\fnm{Thomas} \sur{Lottermoser}}

\author[3]{\fnm{Fabio} \sur{Caruso}}

\author[1]{\fnm{Manfred} \sur{Fiebig}}

\affil[1]{\orgdiv{Department of Materials}, \orgname{ETH Zurich}, \orgaddress{\street{Vladimir-Prelog-Weg 4}, \city{Zurich}, \postcode{8093}, \country{Switzerland}}}

\affil[2]{\orgname{University of W\"{u}rzburg}, \orgdiv{Institute of Physical and Theoretical Chemistry}, \orgaddress{\street{Am Hubland}, \postcode{97074} \city{W\"{u}rzburg}, \country{Germany}}}

\affil[3]{\orgdiv{Institute of Theoretical Physics and Astrophysics}, \orgname{Kiel University}, \orgaddress{\street{Leibnizstra\ss{}e 15}, \city{Kiel}, \postcode{24118}, \country{Germany}}}

\affil[4]{\orgdiv{Université Grenoble Alpes}, \orgname{CNRS, Institut Néel}, \orgaddress{\street{25. av. des Martyrs} \postcode{38042} \city{Grenoble},  \country{France}}}


\abstract{The breakdown of Hooke’s law marks the onset of nonlinear behaviour: when displacements become large, restoring forces weaken and conventional proportionality fails. In quantum materials, intense optical excitation can drive the crystal lattice into a similar regime, where established linear relations between light, electrons, and phonons no longer hold. Sliding ferroelectrics are particularly susceptible, as controlling their polarization requires large interlayer shifts. Displacive excitation of coherent phonons, the principal mechanism for launching structural motion, typically assumes that lattice-driving forces scale linearly with the photo-excited carrier density. Whether this linearity survives at high excitation, however, remains largely unexplored, and its breakdown can fundamentally limit accessible lattice displacements. Here we show that such nonlinear limitations can be surpassed in a sliding ferroelectric by timing, rather than strengthening the optical drive. Time-resolved second-harmonic generation reveals that the interlayer sliding phonon governing ferroelectricity saturates and even diminishes under single-pulse excitation. First-principles calculations attribute this nonlinearity to band-specific electron–phonon coupling that induces competing forces on the lattice. By splitting the optical energy into two well-timed pulses that avoid populating counteracting states, we achieve markedly larger phonon amplitudes at fixed total fluence. The resulting enhanced sliding motion exposes a regime of anharmonic phonon coupling that emerges only far from equilibrium. Our findings show that nonlinear limits in driven solids can be overcome, opening new pathways for steering lattice motion in quantum materials.}

\maketitle

Large-amplitude coherent lattice motion provides a powerful route to controlling functionality in quantum materials~\cite{weiner_femtosecond_1990,dougherty_femtosecond_1992,forst_mode-selective_2015,basini_terahertz_2024,shi_nonresonant_2024,de_la_torre_colloquium_2021,disa_engineering_2021}. Femtosecond light pulses can drive solids far from equilibrium, creating transient structural configurations with emergent ferroic~\cite{nova_metastable_2019,disa_photo-induced_2023} or electronic properties~\cite{rini_control_2007,mitrano_possible_2016,horstmann_coherent_2020,maklar_coherent_2023}. Displacive excitation of coherent phonons (DECP)~\cite{zeiger_theory_1992,stevens_coherent_2002} harnesses photo-excited carriers to launch Raman-active modes and thereby steer atomic motion. Van der Waals sliding ferroelectrics~\cite{fei_ferroelectric_2018,yang_origin_2018,yang_laser-induced_2022,ni_mechanically_2020,zhang_emerging_2025,wu_sliding_2021} are particularly promising for such vibrational control, as their polarization arises from the relative displacement of adjacent layers and remains robust down to the few-layer limit~\cite{zhang_emerging_2025,wu_sliding_2021}. Driving this sliding motion to extreme amplitudes may even enable ferroelectric switching on ultrafast timescales~\cite{yang_laser-induced_2022,yang_origin_2018}. Yet achieving such large displacements within DECP remains challenging, as electronic excitation, damage thresholds, and anharmonic decay at high fluence often limit the attainable vibrational amplitude~\cite{teitelbaum_direct_2018}.\\
\\
Here, we show that splitting an optical pulse into two time-separated excitations overcomes amplitude saturation of the key phonon mode in a van der Waals sliding ferroelectric, enhancing control over ferroelectric polarization and enabling access to nonlinear lattice phenomena. Using time-resolved second-harmonic generation (trSHG)~\cite{shan_giant_2021,satoh_ultrafast_2007} and \textit{ab initio} calculations, we investigate the coupling between interlayer sliding motion and ferroelectric polarization, and trace the nonlinear amplitude limit to the occupation of high-lying electronic states counteracting sliding phonon generation. Staggered sliding-mode stimulation avoids this compensation, restores linearity, and thus amplifies sliding motion beyond the single-pulse limit. By exploiting the enhanced lattice response, we achieve greater control over the ferroelectric polarization and uncover a strong correlation between sliding-mode amplitude, damping, and frequency, revealing a new form of anharmonic coupling mediated by coherent nuclear motion.\\
\\
As a model system, we investigate WTe$_2$, a ferroelectric van der Waals semimetal that has recently attracted attention for its giant, non-saturating magnetoresistance and ultrafast optical switching between topologically distinct states~\cite{sharma_room-temperature_2019,sie_ultrafast_2019,ali_large_2014,li_evidence_2017,guan_manipulating_2021,ji_manipulation_2021,hein_mode-resolved_2020}. In its thermodynamically stable ground state, the Td polymorph of WTe$_2$ (stable for temperatures $T<565\,\text{K}$) adopts a trilayer structure that breaks inversion symmetry perpendicular to the van der Waals planes (see calculated potential energy surface in Fig.\,\ref{fig:1}a and lattice structure in Fig.\,\ref{fig:1}b). An uncompensated interlayer vertical charge transfer gives rise to a ferroelectric polarization oriented along the crystallographic $c$-axis, that is, normal to the layers~\cite{fei_ferroelectric_2018,yang_origin_2018,yang_laser-induced_2022,ni_mechanically_2020}. Ultrafast optical excitation across a broad spectral range (from terahertz to visible) depopulates antibonding orbitals between neighbouring layers~\cite{sie_ultrafast_2019,ji_manipulation_2021,guan_manipulating_2021,aoki_excitation_2022}. This launches an interlayer sliding mode at $f_{\text{s}} = 0.24\,\text{THz}$, which pushes the system towards a high-symmetry configuration on the potential energy surface, eventually restoring inversion symmetry~\cite{sie_ultrafast_2019,ji_manipulation_2021,he_coherent_2016,dai_ultrafast_2015,soranzio_ultrafast_2019,soranzio_strong_2022} (see Fig.\,\ref{fig:1}a and displacement pattern in Fig.\,\ref{fig:1}b).  Such interlayer sliding motion is expected to strongly influence ferroelectric properties on ultrafast time scales, a feature shared broadly among van der Waals sliding ferroelectrics~\cite{yang_laser-induced_2022, zhang_emerging_2025}. We point out that although the sliding mode is often referred to in the literature as a `shear mode', it does not break the symmetry of the elastic tensor and is therefore not a shear mode in the strict sense. Thus, it is essential to carefully distinguish between the two terms.\\
\\
In a first step, we use symmetry considerations to investigate how structural distortions connect the ferroelectric and paraelectric configurations of WTe$_2$ (see Fig.\,\ref{fig:1}b). The polar Td phase belongs to the space group $Pmn2_1$, whereas the nonpolar, ferroelastic 1T' phase crystallizes in the $P2_1/m$ structure. While the interlayer sliding mode is often discussed in the context of the $\text{Td}\rightarrow 1\text{T'}$ transition, it should be noted that the 1T' and Td phases are not related by a group–subgroup relationship~\cite{ding_phase_2021}. Therefore, the sliding mode does not, by itself, provide a symmetry-allowed pathway between them. Instead, we identify the centrosymmetric $Pnma$ structure as the true high-symmetry reference phase, from which both the 1T' and Td polymorphs emerge as symmetry-lowering instabilities. The two relevant distortions are the interlayer sliding mode ($\Gamma_2^-$ irreducible representation), leading to the Td phase, and an intralayer shear mode ($\Gamma_4^+$ irreducible representation), giving rise to the 1T' phase. The structural pathway connecting the Td and 1T' phases thus requires both interlayer sliding and intralayer shearing, with the lowest-energy route bypassing the energetically unfavourable $Pnma$ structure through the simultaneous excitation of both modes (Fig.\,\ref{fig:1}a; see Methods for details on the underlying density functional theory calculations).\\
\\
In order to relate the free-energy landscape $F$ to the vibrational spectrum, we perform an anharmonic phonon expansion around the minimum energy structure (Td phase), yielding
\begin{align}
F = F_{\mathrm{Td}} + \omega_{A_1}^2 Q_{A_1}^2 + \omega_{B_1}^2 Q_{B_1}^2 + \gamma Q_{A_1} Q_{B_1}^2,\label{eq:NLcoupling}
\end{align}
where $ Q_{A_1}\sim (P-P_z)$ represents the amplitude of the interlayer-sliding phonon, and $Q_{B_1}$ the intralayer-shear phonon. The corresponding mode frequencies are $\omega_{A_1}/2\pi := f_{\text{s}}= 0.24\,\text{THz}$ and $\omega_{B_1}/2\pi = 2.7\,\text{THz}$. Thus, interlayer sliding couples directly to the ferroelectric polarization $P_z$, whereas intralayer shearing couples only indirectly to $P_z$ through nonlinear phonon–phonon interactions, parametrized by the last term in Eq.\,\ref{eq:NLcoupling}.

\subsection*{Sliding mode amplitude limitation}
SHG, owing to its inherent sensitivity to inversion-symmetry breaking, is an ideal probe of structural dynamics governing ferroelectric order~\cite{denev_probing_2011}. In our trSHG experiments, we use 130-fs pump pulses ($\hbar\omega_{\text{p}} = 0.82\,\text{eV}$) to trigger structural dynamics, and detect the SHG response to the probe pulses ($\hbar\omega_{\text{pr}} = 1.55\,\text{eV}$) at variable time delays $\Delta t_{\text{p-pr}}$ and optical polarizations (see Methods and Supplementary Fig.\,S5). Tracking the transient symmetry changes in Td-WTe$_2$ following single-pulse optical excitation reveals an initial drop in SHG intensity within 500\,fs, concomitant with prominent oscillations at the interlayer-sliding-mode frequency $f_{\text{s}}$ (see Figs.\,\ref{fig:1}d).\\
\\
Maximizing the sliding-mode amplitude provides greater leverage over the ferroelectric polarization. To explore the limits of the linear coupling between electronic excitation and the sliding-mode amplitude, we perform fluence-dependent trSHG measurements (see Fig.\,\ref{fig:1}c,d). Increasing the absorbed fluence from $0.4-6.5\,\text{mJ}\,\text{cm}^{-2}$ progressively shifts the equilibrium position $Q_0$ towards the centrosymmetric $Pnma$ configuration (compare Fig.\,\ref{fig:1}a) and leads to a pronounced softening of the sliding mode. This is reflected in a long-lived signal suppression and fluence-dependent frequency shifts, respectively (Fig.\,\ref{fig:1}f). Most notably, the sliding-mode amplitude initially increases with fluence, then reaches a maximum around an absorbed fluence of $F_{\text{sat}} \approx 1.6\,\text{mJ}\,\text{cm}^{-2}$, and even declines at higher excitation levels (see red trace in Fig.\,\ref{fig:1}e). This counterintuitive reduction imposes a fundamental limit on the extent to which coherent lattice motion can manipulate ferroelectricity. Together with transient reflectivity and time-resolved X-ray diffraction measurements on WTe$_2$~\cite{soranzio_strong_2022} and MoTe$_2$~\cite{fukuda_coherent_2024}, our results suggest that such saturation behaviour is a general characteristic of these systems.\\
\\
Understanding the fluence-dependent behaviour of the sliding-mode amplitude requires going beyond the conventional, linear DECP model, which assumes a linear relationship between excited-carrier density and coherent-phonon amplitude~\cite{zeiger_theory_1992}. Instead, we account for the contributions of individual electronic states at different energies to the effective electron–phonon coupling at the sliding-mode frequency. To this end, we perform \textit{ab initio} calculations~\cite{emeis_coherent_2025,pan_origin_2025} to determine the driving force on the sliding mode as a function of the optically-induced transient electronic temperature (see Methods for details).\\
\\
Figure\,\ref{fig:2}a displays the state-resolved driving forces for the sliding mode at electronic temperatures $T_{\text{elec}}=2000\,\text{K}$ and $T_{\text{elec}}=4000\,\text{K}$, that is, above and below the amplitude-limiting threshold. The contributions of individual states are weighted by their occupation, governed by changes $\Delta f_{\text{elec}}$ in the Fermi–Dirac distribution at the transient electronic temperature following optical excitation. At low excitation laser fluence, that is, low electronic temperature, changes in occupation of electron- and hole-like states near the Fermi energy $E_{\text{F}}$ generate net positive sliding forces. With increasing fluence, higher-energy states become populated, which instead counteract sliding-mode excitation. This behaviour is not captured by linear DECP theory and requires the state-specific \textit{ab-initio} approach introduced here.\\
\\
In a real-space picture~\cite{nicholson_beyond_2018}, the generation of photo-excited electrons and holes near $E_{\text{F}}$ induces a spatial modulation of the charge density within the unit cell. Through Coulomb interactions, this transient distribution drives nuclear displacements aligned with the sliding mode pattern. In contrast, population of high-energy states produces a more homogeneous charge density, reducing net Coulomb forces on the nuclei and thereby suppressing sliding-mode-related lattice motion (see Fig.\,\ref{fig:2}b). The saturation behaviour is specific to the sliding mode, whereas higher-frequency modes exhibit a continuous, monotonic increase in amplitude. Comparison between experimental data and the \textit{ab initio} model reveals close agreement in the fluence-dependence of the sliding-mode amplitude (Fig.\,\ref{fig:2}b). More specifically, our calculations predict a saturation of the sliding-mode amplitude at an electronic temperature of $T_{\text{sat}}=3100\,\text{K}$. In the experiment, this corresponds to an absorbed fluence of $F_{\text{sat}}=1.6\,\text{mJ}\,\text{cm}^{-2}$, consistent with the experimental value. Beyond this fluence, both experiment and theory show a decline in sliding-mode amplitude.

\subsection*{Coherent control and amplification of sliding motion}
The amplitude limitation restricts coherent vibrational manipulation of the material’s ferroelectric~\cite{fei_ferroelectric_2018,sharma_room-temperature_2019,yang_laser-induced_2022} and topological~\cite{sie_ultrafast_2019,guan_manipulating_2021} properties to low fluences. This motivates the search for control schemes that beat the limit posed by electronic excitation and reach hitherto inaccessible phonon amplitudes. Specifically, we aim to overcome the sliding-mode amplitude limitation by a double-pulse excitation scheme (Fig.\,\ref{fig:3}a)~\cite{akei_role_2025,hase_femtosecond_2015,horstmann_coherent_2020,horstmann_structural_2024,maklar_coherent_2023}. Repeated in-phase excitation of coherent phonons by optical pulse trains has been shown to selectively amplify targeted vibrational modes to high amplitudes~\cite{weiner_femtosecond_1991,nelson_prospects_1991}. Such `cold' driving schemes rely on vibrational coherence persisting beyond the electronic excitation. Our single-pulse excitation experiments confirm that the sliding mode in WTe$_2$ satisfies this condition, supported by earlier studies~\cite{soranzio_ultrafast_2019,soranzio_strong_2022,sie_ultrafast_2019,ji_manipulation_2021,he_coherent_2016,dai_ultrafast_2015}.\\
\\
We therefore split the pump pulses in two and record trSHG traces as a function of the mutual delay $\Delta t_{\text{p-p}}$ between the two pump pulses and the double-pump-probe delay $\Delta t_{\text{p-pr}}$ (see Fig.\,\ref{fig:3}b). Strong modulations of the vibrational amplitude for different double-pulse delays $\Delta t_{\text{p-p}}$ demonstrate coherent control over the sliding-mode amplitude $A_{\text{s}}$ in the excited state (see selected SHG traces in Fig.\,\ref{fig:3}c), corroborated by \textit{ab initio} calculations (Fig.\,\ref{fig:3}d). Comparing double-pulse SHG traces for $\Delta t_{\text{p-p}}=0$, where both pump pulses act as a single excitation, and $\Delta t_{\text{p-p}}=2T_{\text{s}}$ ($T_{\text{s}}=2\pi/f_{\text{s}}$), we find that, for equal total optical energy, sequential optical excitation yields up to 50\,\% higher sliding amplitudes (see Fig.\,\ref{fig:3}e). This key result is visualized in Fig.\,\ref{fig:3}g, which directly compares the yields of single- and double-pulse excitation and shows that sequential excitation extends the linear relation between optical energy and sliding amplitude in the targeted nonequilibrium state.\\
\\
Analyzing the sliding-mode amplitude after the second excitation as a function of $\Delta t_{\text{p-p}}$ further reveals a strong suppression for overlapping pump pulses compared to the delay-dependent amplitude expected in coherent control experiments (see red-shaded area near $\Delta t_{\text{p-p}}=0$ in Fig.\,\ref{fig:3}f). The suppression persists for inter-pulse delays up to $\sim$2\,ps (compare violet trace in Fig.\,\ref{fig:3}f). This timescale agrees with the carrier lifetimes determined by time- and angle-resolved photoemission spectroscopy~\cite{hein_mode-resolved_2020}, supporting our hypothesis that excitation of charge carriers to high energies above the Fermi level counteracts sliding-mode generation. We note that saturable absorption in WTe$_2$ on a few-ps timescale~\cite{gao_broadband_2018} cannot account for this behaviour as the magnitude of that effect appears insufficient to explain the pronounced sliding mode saturation and, in particular, its subsequent decline.

\subsection*{Large-amplitude coherent vibrational spectroscopy}
In a next step, we exploit the fact that we can beat amplitude saturation by well-timed sequential excitation to enable large-amplitude coherent vibrational spectroscopy in the excited state. In particular, double-pulse excitation not only facilitates enhanced vibrational amplitudes but also allows precise control over the amplitude while keeping other key degrees of freedom, such as electronic excitation and temperature, constant~\cite{scharf_coherent_2025,ravnik_time-domain_2021,cheng_coherent_2017}. This makes it possible to isolate correlations between the coherent sliding-mode amplitude on the one hand, and its damping and frequency on the other hand (see Fig.\,\ref{fig:4}a)~\cite{cheng_coherent_2017}.\\
\\
Applying this scheme, we observe a strong correlation between the sliding mode’s coherent amplitude $A_{\text{s}}$ and its damping constant $\gamma_{\text{s}}=\tau_{\text{s}}^{-1}$ across the full amplitude range (Fig.\,\ref{fig:4}a,b). This behaviour indicates that the damping is directly mediated by coherent nuclear motion. Crucially, such a decay channel is absent in equilibrium crystals, where nuclear wave packets remain centered at their equilibrium positions irrespective of temperature. Amplitude-dependent damping traces recorded at different combined fluences reveal a strikingly linear relation, characterized by the slope $(\Delta\gamma/\Delta A)_{\text{s}}$, with larger coherent amplitudes producing proportionally stronger damping (Fig.\,\ref{fig:4}b). Notably, $(\Delta\gamma/\Delta A)_{\text{s}}$ itself increases with combined fluence, such that for a given initial sliding amplitude, higher overall excitation enhances the damping.\\
\\
To identify the microscopic origin of this amplitude- and fluence-dependent damping, we consider a Klemens process ~\cite{dyson_role_2013,klemens_anharmonic_1966,hejda_hot-electron_1993} as the most probable decay mechanism for the sliding mode. As the lowest-energy optical phonon, the sliding mode can decay into two lower-energy acoustic phonons with opposite momentum (Fig.\,\ref{fig:4}c). Stimulated phonon decay has been predicted to strongly enhance optical-phonon damping under nonequilibrium conditions, with the decay rate of the optical phonon proportional to the populations of the optical and acoustic modes.~\cite{romanov_effects_1999}\\
\\
Consistent with this picture, the two experimental trends established above naturally emerge from stimulated decay of the sliding mode into acoustic modes. First, the damping increases with the sliding-mode population, or equivalently its coherent amplitude. Second, the damping grows with combined fluence, as higher overall excitation enhances scattering into low-energy acoustic branches. Thus, stimulated phonon decay provides a microscopic rationale for the observed amplitude- and fluence-dependent damping.\\
\\
The sliding-mode frequency, on the other hand, exhibits a clear anti-correlation with the sliding amplitude (compare red and brown traces in Fig.\,\ref{fig:4}a): as the amplitude increases, the frequency softens. Since both the lattice temperature and the level of electronic excitation are fixed in double-pulse experiments, a significant contribution from electronic softening, as reported in other semimetals~\cite{murray_effect_2005}, can be excluded. We therefore attribute the observed amplitude-dependent frequency shift to intrinsic lattice anharmonicity, explored by large-amplitude coherent nuclear motion.\\
\\
In pump–probe experiments, even such clear linear relations between coherent phonon amplitude, damping, and frequency are difficult to extract unambiguously, as variations in phonon amplitude are intertwined with electronic excitation. Double-pulse excitation overcomes this limitation by enabling independent control of the phonon amplitude within a targeted excited state.

\subsection*{Discussion and summary}
In summary, we have demonstrated that redistributing the energy of a single ultrafast excitation into two pulses overcomes the amplitude limitations of key vibrational modes, enhancing control over ferroic order and revealing nonequilibrium lattice dynamics. Our results show that the linear relationship between excited-carrier density and coherent vibrational amplitude assumed in DECP~\cite{zeiger_theory_1992} is an oversimplification, highlighting the potential of sequential excitation to surpass the resulting nonlinear limits and access large-amplitude dynamics. Such strategies are particularly advantageous in materials where decisive modes are exclusively Raman-active and coupled IR-active modes possess only weak dipole moments, rendering nonlinear phononic driving~\cite{forst_mode-selective_2015,nova_metastable_2019,disa_photo-induced_2023} ineffective.\\
\\
Owing to the shared structural motifs of van der Waals materials and the universality of DECP, we expect the principles demonstrated here for WTe$_2$ to be broadly applicable to other systems and their vibrational degrees of freedom, particularly other sliding ferroelectrics. In this context, large-amplitude vibrational spectroscopy enabled by sequential excitation provides a powerful means to investigate and tailor phonon-electron and phonon-phonon interactions far from equilibrium~\cite{cheng_coherent_2017,ravnik_time-domain_2021}. We envision that leveraging vibrational coherences near topologically trivial and paraelectric configurations opens new avenues for coherent control of both topology and ferroelectricity in functional materials.

\newpage

\section*{Methods}\label{secA1}

\begin{figure}[ht!]
    \centering
    \makebox[\textwidth][c]{
    \includegraphics[width=1.4\linewidth]{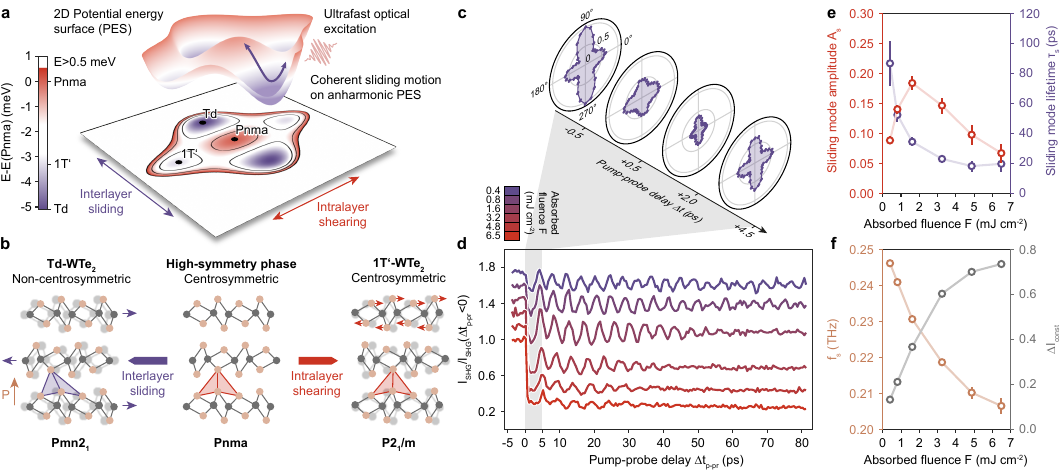}}
    \caption{\textbf{$\mid$ Ultrafast structural dynamics in ferroelectric Td-WTe$_2$.} \textbf{a}, Calculated two-dimensional potential energy surface of WTe$_2$ spanned by interlayer sliding and intralayer shear displacements. \textbf{b}, Structural configurations of WTe$_2$ and lattice modes connecting the Td and 1T' polymorphs to the paraelectric high-symmetry phase. Dark grey circles, tungsten atoms; light brown circles, tellurium atoms; grey background structures denote the atomic configuration of the high-symmetry paraelectric phase, included for comparison. \textbf{c}, Polarization-dependent SHG intensity at selected pump–probe delays $\Delta t_{\text{p-pr}}$. The incident polarization of the fundamental is varied, with the analyzer fixed in the horizontal orientation. \textbf{d}, Normalized, delay-dependent SHG efficiency (polarizer: vertical; analyzer: horizontal) as a function of absorbed laser fluence $F$. Traces are vertically offset for clarity. \textbf{e}, Sliding-mode amplitude $A_\text{s}$ and lifetime $\tau_\text{s}$ as a function of absorbed fluence $F$. \textbf{f}, Sliding-mode frequency and quasi-static suppression of SHG intensity as functions of absorbed fluence $F$.}
    \label{fig:1}
\end{figure}

\begin{figure}[ht!]
    \centering
    \includegraphics[width=0.7\linewidth]{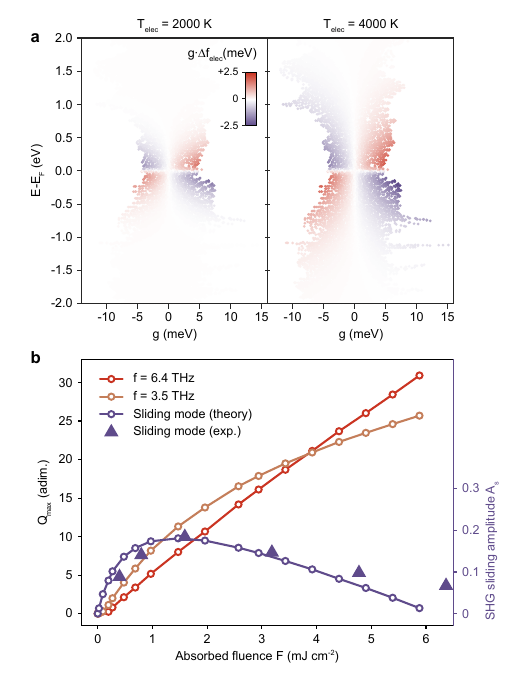}
    \caption{\textbf{$\mid$ Ab initio modeling of sliding-amplitude saturation.} \textbf{a}, State-resolved electron-phonon coupling strength $g$ at the sliding-mode frequency at two electronic temperatures $T_{\text{elec}}=2000\,\text{K}$ and $T_{\text{elec}}=4000\,\text{K}$. Red (blue) indicates positive (negative) contributions to the sliding mode excitation with more saturated colors corresponding to stronger coupling. Along the energy axis, the coupling strength $g$ is weighted by the change in the Fermi–Dirac distribution $\Delta f_{\text{elec}}$ at the respective temperature. \textbf{b}, Calculated maximum amplitudes $Q_{\text{max}}$ of the sliding mode and of two higher-frequency modes of Td-WTe$_2$ as a function of absorbed fluence $F$. Triangular data points represent the sliding-mode amplitudes determined in our experiments for comparison.}
    \label{fig:2}
\end{figure}

\begin{figure}[ht!]
    \centering
    \makebox[\textwidth][c]{
    \includegraphics[width=1.4\linewidth]{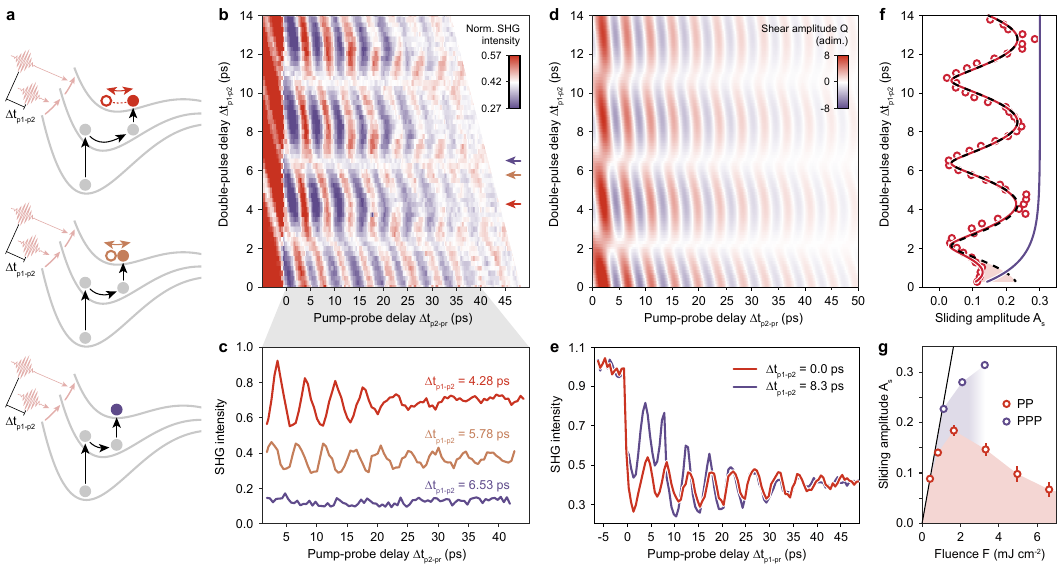}}
    \caption{\textbf{$\mid$ Coherent control and amplification of sliding motion.} \textbf{a}, Schematic of amplitude control of displacively excited coherent phonons using double-pulse excitation. \textbf{b}, Normalized SHG intensity as a function of pump–probe delay $\Delta t_{\text{p-pr}}$ and double-pulse delay $\Delta t_{\text{p1-p2}}$ for a combined absorbed fluence of $2.1\,\text{mJ}\,\text{cm}^{-2}$. Red, yellow, and violet arrows indicate the double-pulse delays corresponding to the pump–probe traces shown in (c). \textbf{c}, Pump–probe SHG traces at selected double-pulse delays, demonstrating coherent control of the sliding-mode amplitude. \textbf{d}, Simulated sliding-mode amplitude $Q$ as a function of $\Delta t_{\text{p-pr}}$ and $\Delta t_{\text{p1-p2}}$, assuming an amplitude-dependent sliding frequency. \textbf{e}, Comparison of pump–probe SHG traces for $\Delta t_{\text{p1-p2}} = 0\,\text{ps}$ and $\Delta t_{\text{p1-p2}} = 8.3\,\text{ps}$. \textbf{f}, Sliding-mode amplitude after the second optical excitation as a function of $\Delta t_{\text{p1-p2}}$. Dashed black line, expected behaviour in coherent control experiments without amplitude saturation (see Methods); red line, fitted model incorporating amplitude suppression at short double-pulse delays; violet line, transient suppression of the sliding mode extracted from fits to the experimental data (see Methods for details). \textbf{g}, Comparison of vibrational amplitudes extracted from pump–probe (PP) and pump–pump–probe (PPP) measurements. Black line, estimated vibrational amplitude assuming a linear dependence of $A_\text{s}$ on the excited-carrier density.}
    \label{fig:3}
\end{figure}

\begin{figure}[ht!]
    \centering
    \makebox[\textwidth][c]{
    \includegraphics[width=1.4\linewidth]{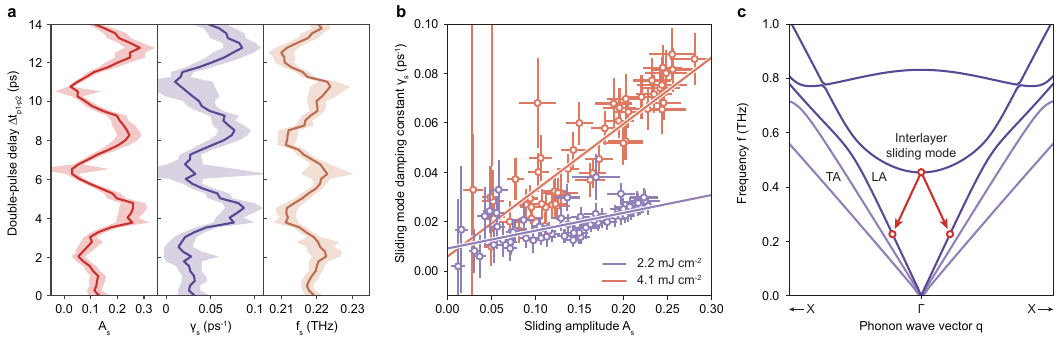}}
    \caption{\textbf{$\mid$ High-amplitude vibrational spectroscopy in targeted excited states.} \textbf{a}, Sliding-mode amplitude $A_\text{s}$, damping $\gamma_{\text{s}}=\tau_{\text{s}}^{-1}$, and frequency $f_{\text{s}}$ as a function of the double-pulse delay $\Delta t_{\text{p1-p2}}$ for a combined absorbed fluence of $2.1\,\text{mJ}\,\text{cm}^{-2}$. Light-colored shades,  $1\sigma$-confidence intervals of the fits used to extract parameters from the two-dimensional dataset shown in Fig.\,\ref{fig:3}b. \textbf{b}, Sliding-mode damping constant $\gamma_{\text{s}}$ as a function of the sliding-mode amplitude $A_{\text{s}}$ for two different combined fluences in double-pulse experiments. Solid red and violet lines, linear models fitted to the data to determine  $(\Delta\gamma/\Delta A)_{\text{s}}$; errorbars, $1\sigma$-confidence intervals of the fits used to extract $A_{\text{s}}$ and $\gamma_{\text{s}}$. \textbf{c}, Low-energy phonon band structure of WTe$_2$. Red arrows indicate an exemplary pathway for the Klemens-like decay of the sliding phonon into two acoustic phonons of opposite momentum.}
    \label{fig:4}
\end{figure}

\textbf{Sample preparation.} WTe$_2$ single crystals (commercially ordered from HQ Graphene) were exfoliated to remove the oxidized surface layers and immediately mounted in an optical-microscopy cryostat. The sample space was constantly purged with nitrogen gas to minimize oxidization of the pristine surface layers. SHG and oscillatory signals from the sample showed less than $2\,\%$ degradation over several weeks. All experiments were carried out at room temperature.\\
\\
\textbf{Experimental setup.} In single-pulse trSHG measurements, structural dynamics are initiated by focusing pump pulses (center wavelength $\lambda_{\text{p}} = 1.5\,\mu\text{m}$, photon energy $\hbar\omega_{\text{p}} = 0.83\,\text{eV}$, pulse duration $\Delta\tau_{\text{p}} = 130\,\text{fs}$, absorbed fluence $F_{\text{p}} = 0.4–6.5\,\text{mJ}\,\text{cm}^{-2}$, polarization along the $b$-axis of the WTe$_2$ crystal, repetition rate $f_{\text{rep}} = 1\,\text{kHz}$) onto the sample surface under normal incidence using a plano-convex lens ($f_{\text{pump}} = 100\,\text{mm}$). The resulting dynamics are probed by monitoring the SHG signal produced by the probe pulses (center wavelength $\lambda_{\text{pr}} = 0.8\,\mu\text{m}$, photon energy $\hbar\omega_{\text{pr}} = 1.55\,\text{eV}$, pulse duration $\Delta\tau_{\text{p}} = 120\,\text{fs}$, fluence $F_{\text{p}} = 2.8\,\text{mJ}\,\text{cm}^{-2}$, repetition rate $f_{\text{rep}} = 1\,\text{kHz}$) as a function of the time delay $\Delta t_{\text{p-pr}}$ set by a motorized delay stage, and for different combinations of incident and detected probe polarizations. Probe pulses are focused onto the sample in the $ac$-plane and at a $45^\circ$ angle to the surface normal using a plano-convex lens ($f_{\text{pump}} = 75\,\text{mm}$). For the measurements shown in Figs.\,\ref{fig:1}-\,\ref{fig:4}, SHG signals are acquired in a cross-polarized configuration, with the incident (detected) probe polarization aligned parallel (perpendicular) to the $ac$-plane of the crystal.\\
\\
Parasitic SHG signals originating from optical elements in the probe path are suppressed using a long-pass coloured-glas filter (Schott RG630, $2\,\text{mm}$). After reflection from the sample, SHG and fundamental beams are separated using a short-pass filter (Schott BG39, $2\,\text{mm}$) and a band-pass filter (center wavelength $400\,\text{nm}$, bandwidth $25\,\text{nm}$). The SHG signal is collected by a plano-convex lens (focal length $f = 75\,\text{mm}$), detected using a photomultiplier tube, and processed with electronic amplification and time-domain filtering via a boxcar integrator to enhance the signal-to-noise ratio. Each data point is averaged over 5000 pulses ($t_{\text{int}}=5$\,s). The temporal overlap between pump and probe pulses (time zero) is determined by sum-frequency generation in a beta-barium-borate (BBO) crystal placed at the sample position. To ensure homogeneous excitation of the probed region, the pump spot size on the sample ($280 \times 280\,\mu\text{m}^2$) is chosen significantly larger than the probe spot size ($80 \times 40\,\mu\text{m}^2$).\ \\
\\
For double-pulse trSHG measurements, pump-pulse pairs are generated using a Mach–Zehnder-type interferometer and focused colinearly onto the sample. The relative delay $\Delta t_{\text{p1-p2}}$ between pulses p$_1$ and p$_2$ is controlled by a motorized delay stage in one interferometer arm. The pulse energies and polarizations of p$_1$ and p$_2$ are adjusted independently using neutral-density glass filters and half-wave plates placed in each arm. Temporal overlap ($\Delta t_{\text{p1-p2}} = 0$) is determined by measuring the SHG signal from the sample as a function of $\Delta t_{\text{p1-p2}}$ for a fixed pump-probe delay ($\Delta t_{\text{p1-pr}} = 100\,\text{ps}$) with both pump pulses polarized vertically relative to the $ac$-plane of the sample (see Fig.\,S1).\\
\\
To suppress coherent artifacts caused by interference of the two pump pulses near $\Delta t_{\text{p1-p2}} = 0$, all actual measurements are performed with pump pulses p$_1$ and p$_2$ in a cross-polarized configuration. Cross-correlation scans of the SHG signal as a function of the relative polarization angle $\alpha_{\text{p1,p2}}$ confirm the absence of measurable coherent artifacts at $\alpha_{\text{p1,p2}} = 90^{\circ}$ (see Fig.\,S1). During double-pulse measurements, the SHG signal is recorded as a function of $\Delta t_{\text{p1-p2}}$ at each time point along the pump–probe delay, resulting in the two-dimensional data sets shown in Figs. 2c and 3a.\\
\\
\textbf{Analysis of the pump-probe data.} For the time-resolved polarizer scan in Fig.\,\ref{fig:1}c, the SHG intensity $I_{\text{SHG}}(\Delta t_{\text{p-pr}}\,\phi)$ is normalized to the average pre-pump value $I_{\text{SHG}}(\Delta t_{\text{p-pr}}<0,\phi=90\,^{\circ})$. For the fluence- and delay-dependent traces in Fig.\,\ref{fig:1}e, the SHG intensity $I_{\text{SHG}}(\Delta t_{\text{p-pr}})$ is normalized to the average pre-pump value $I_{\text{SHG}}(\Delta t_{\text{p-pr}}<0)$. For clarity, traces corresponding to different fluences are vertically offset by a constant. To extract the fluence dependent amplitudes, decay constants, sliding-mode frequencies and long-lived signal suppression (see Fig. 1e and 1f), the model function
\begin{align}
    I_{\text{SHG}}(\Delta t_{\text{p-pr}}) &= \left[\Delta t_{\text{p-pr}}\leq 0\right] + \left(1-\Delta I_{\text{PT}}-\Delta I_{\text{SP}}-\Delta I_{\text{fast}}^{\text{exc}}\cdot\Delta I_{\text{fast}}^{\text{decay}}\right)
\end{align}
with the individual terms
\begin{align*}
    \Delta I_{\text{PT}}(\Delta t_{\text{p-pr}}) &= \Delta I_{\text{PT}}^{0}\cdot\left(1-\text{exp}\left(-\frac{\Delta t_{\text{p-pr}}}{\tau_{\text{PT}}}\right)\right),\\
    \Delta I_{\text{SP}}(\Delta t_{\text{p-pr}}) &= -\Delta I_{\text{SP}}^{0}\cdot\text{cos}\left(\frac{2\pi\Delta t_{\text{p-pr}}}{T_{\text{sliding}}}+\phi_{\text{sliding}}\right)\cdot\text{exp}\left(-\frac{-\Delta t_{\text{p-pr}}}{\tau_{\text{sliding}}}\right),\\
    \Delta I_{\text{fast}}^{\text{exc}}(\Delta t_{\text{p-pr}}) &= \Delta I_{\text{fast}}^{\text{exc},0}\cdot\left(1-\text{exp}\left(-\frac{\Delta t_{\text{p-pr}}}{\tau_{\text{fast}}^{\text{exc}}}\right)\right),\text{ and}\\
    \Delta I_{\text{fast}}^{\text{decay}}(\Delta t_{\text{p-pr}}) &= \Delta I_{\text{fast}}^{\text{decay},0}\cdot\left(\text{exp}\left(-\frac{\Delta t_{\text{p-pr}}}{\tau_{\text{fast}}^{\text{decay}}}\right)\right)
\end{align*}
is fitted to the delay-dependent traces in Fig.\,\ref{fig:1}d. Here, $\Delta t_{\text{p-pr}}$ is the pump-probe delay, $\Delta I_{\text{PT}}$ corresponds to the change in the SHG signal due to the shift of the equilibrium position after photo-excitation (see also Ref.~\cite{sie_ultrafast_2019}  for details), $\Delta I_{\text{SP}}$ is the modulation of the SHG intensity due to coherent sliding motion, and $\Delta I_{\text{fast}}^{\text{exc}}$ or $\Delta I_{\text{fast}}^{\text{decay}}$, respectively, model the fast initial suppression of the SHG signal. Parameters $\tau_{\text{PT}}$ and $\tau_{\text{sliding}}$ correspond to the characteristic time scale of the structural shift or the sliding mode coherence time, respectively. Furthermore, $\tau_{\text{fast}}^{\text{exc}}$ and $\tau_{\text{fast}}^{\text{decay}}$ are the time constants for the excitation and subsequent decay of the fast initial signal suppression observed in the $\chi_{zyy}$ component. $T_{\text{sliding}}$ is the sliding-mode period.\\
\\
\textbf{Analysis of the pump-pump-probe data.} To determine the sliding-mode amplitude $A_{\text{s}}$, frequency $f_{\text{s}}$, phase $\phi_{\text{s}}$, and damping $\gamma_{\text{s}}=\tau_{\text{s}}^{-1}$, as well as the quasi-dc signal suppression $\Delta I_{\text{const}}$ in double-pulse experiments, we fit the model
\begin{align}
I_{\text{SHG}}(\Delta t_{\text{p-pr}}) = A_{\text{s}}\cdot\text{exp}\left(-\frac{\Delta t_{\text{p-pr}}}{\tau_{\text{s}}}\right)\cdot\text{cos}(2\pi f_{\text{s}}\Delta t_{\text{p-pr}}+\phi_{\text{s}})+\Delta I_{\text{const}}
\end{align}
to the trSHG trace after the second excitation for each double-pulse delay $\Delta t_{\text{p-p}}$. The results are shown in Fig.\,\ref{fig:4}a. The $\Delta t_{\text{p1-p2}}$-dependent sliding-mode amplitude $A_{\text{s}}$ in Fig.\,\ref{fig:3}f, on the other hand, is fitted by the model
\begin{align}
A_{\text{s}}(\Delta t_{\text{p1-p2}}) = A_{\text{s}}^{\text{osc}}(\Delta t_{\text{p1-p2}})\cdot \left(1-A_{\text{e}}^{\text{damp}}\cdot\text{exp}\left(-\frac{\Delta t_{\text{p1-p2}}}{\tau_{\text{e}}^{\text{damp}}}\right)\right),
\end{align}
where
\begin{align}
A_{\text{s}}^{\text{osc}}(\Delta t_{\text{p1-p2}}) = A_{\text{s}}\sqrt{1+2\zeta\cdot(2\pi f_{\text{s}}\Delta t_{\text{p1-p2}})+\zeta^2}.
\end{align}
Here, $A_{\text{e}}^{\text{damp}}$ is the leverage of electronic excitation for the damping of the sliding-mode amplitude and $\tau_{\text{e}}^{\text{damp}}$ the electronic relaxation time. Furthermore, the quotient $(\Delta\gamma_{\text{s}}/\Delta A_{\text{s}})(T)$ is determined by fitting the linear model 
\begin{align}
\gamma_{\text{s}}(A_{\text{s}}) = A_{\text{s}}^0+(\Delta\gamma_{\text{s}}/\Delta A_{\text{s}})\cdot A_{\text{s}}
\end{align}
to the two datasets presented in Fig.\,\ref{fig:4}b.\\
\\
\textbf{Fitting of the polarization-dependent trSHG data.} 
To fit the SHG polarimetry data shown in Fig.\,\ref{fig:1}c and Fig.\,S2, we consider the symmetry of the Td polytype. In the orthorhombic Td phase, WTe$_2$ belongs to the $mm2$ point group. We consider only electric-dipole–allowed SHG, described by the odd-parity, polar (i-type) second-order nonlinear susceptibility tensor $\chi^{(2)}$. According to Ref.\,\cite{birss_macroscopic_1963}, the non-vanishing components of $\chi^{(2)}$ for this symmetry are $\chi^{(2)}_{zzz}$, $\chi^{(2)}_{xxz}$, $\chi^{(2)}_{xzx}$, $\chi^{(2)}_{zxx}$, $\chi^{(2)}_{yyz}$, $\chi^{(2)}_{yzy}$, and $\chi^{(2)}_{zyy}$. The incident electric field is given by
\begin{align}
    \bm E = E_0\begin{pmatrix}
        \cos(\varphi)/\sqrt{2}\\
        \sin(\varphi)\\
        \cos(\varphi)/\sqrt{2}
    \end{pmatrix},
\end{align}
where $\varphi = 0^\circ$ corresponds to p-polarized and $\varphi = 90^\circ$ to s-polarized light relative to the plane of incidence. While SHG is collected in reflection at a $45^\circ$ angle with respect to the surface normal, for simplicity we perform the symmetry analysis in a transmission geometry, which yields equivalent selection rules. The induced second order nonlinear polarization $\bm{P}(2\omega)$ is thus given by
\begin{align}
    \bm{P}(2\omega) &= \begin{pmatrix}
        2\chi_{xxz}E_xE_z\\
        2\chi_{yyz}E_yE_z\\
        \chi_{zxx}E_x^2 + \chi_{zyy}E_y^2 +\chi_{zzz}E_z^2
    \end{pmatrix}\\
    &=E_0^2\begin{pmatrix}
        \chi_{xxz}\cos^2(\varphi)\\
        \sqrt{2}\chi_{yyz}\sin(\varphi)\cos(\varphi)\\
        \frac12(\chi_{zzz}+\chi_{zxx})\cos^2(\varphi) + \chi_{zyy}\sin^2(\varphi)
    \end{pmatrix}.
\end{align}
Using trigonometric identities we yield
\begin{align}
    \bm{P}(2\omega) =E_0^2\begin{pmatrix}
        \chi_{xxz}(1+\cos(2\varphi))/2\\
        \chi_{yyz}\sin(2\varphi)/\sqrt{2}\\
        \frac{1}{4}(\chi_{zzz}+\chi_{zxx})(1+\cos(2\varphi))+\frac{1}{2}\chi_{zyy}(1-\cos(2\varphi))
    \end{pmatrix}.
\end{align}
To calculate the SHG signal after the analyzer, we project the nonlinear polarization $\mathbf{P}(2\omega)$ onto the s- and p-polarized components using the unit vectors $\hat{\mathbf{v}} = (0,1,0)$ and $\hat{\mathbf{h}} = \frac{1}{\sqrt{2}}(1,0,1)$, respectively:
\begin{align}
    I_{\text{h}} \sim |\hat{\bm h} \cdot \bm P|^2 &= |\frac{1}{\sqrt{2}}(P_x+P_z)|^2\\
    &= \frac{E_0^4}{2}|\chi_\text{eff}\cos^2(\varphi)+\chi_{zyy}\sin^2(\varphi)|^2;\\
    I_{\text{v}}\sim |P_y|^2 &= 2E_0^4|\chi_{yyz}\sin(\varphi)\cos(\varphi)|^2,
\end{align}
where $\chi_\text{eff}=\left(\chi_{xxz}+1/2(\chi_{zzz}+\chi_{zxx})\right)$. For $I_h$, we additionally assume a phase difference $\theta$ between $\chi_\text{eff}$ and $\chi_{zyy}$, leading to
\begin{align}
I_h \sim |\chi_\text{eff}\cos^2(\varphi)+e^{i\theta}\chi_{zyy}\sin^2(\varphi)|^2.
\end{align}
While all accessible tensor elements oscillate coherently at the same frequency, a strong initial suppression is observed exclusively in the vertical lobe associated with the $\chi_{zyy}$ component (compare frames at $-0.5\,\text{ps}$ and $+0.5\,\text{ps}$ in Fig.\,\ref{fig:1}c; for further details, see Fig.\,S2). We speculate that this anisotropic response results from an additional structural distortion along the $c$-axis, as suggested by density functional theory calculations under optical excitation \cite{yang_laser-induced_2022}.\\
\\
\textbf{Estimation of the transient lattice-temperature increase.} The transient temperature increase $\Delta T$ within the probed sample volume induced by the pump pulses in the pump–pump–probe experiments is estimated as
\begin{align}
    \Delta T\approx \frac{F(1-R)}{\rho\,c_{\text{p}}\,d},
\end{align}
where $F$ is the incident laser fluence, $R=0.49$ the reflectivity coefficient of WTe$_2$ at the pump wavelength and normal incidence, $\rho=9.43\cdot10^3\,\text{kg}\,\text{m}^{-3}$ is the material density, $c_{\text{p}}=1.77\cdot10^2\,\text{J}\,\text{kg}^{-1}\,\text{K}^{-1}$ is the specific heat capacity at $300\,\text{K}$, $d=-\ln(1/\text{e})/\alpha=48\,\text{nm}$ is the approximate optical penetration depth with $\alpha=2.07\cdot10^5\,\text{cm}^{-1}$ denoting the absorption coefficient at the pump wavelength. The penetration depth of the probe pulses ($\lambda_{\text{pr}}=800\,\text{nm}$, $d_{\text{pr}}=35\,\text{nm}$) is smaller than that of the pump pulse ($\lambda_{\text{p}}=1500\,\text{nm}$, $d_{\text{p}}=48\,\text{nm}$). We therefore neglect any effects arising from inhomogeneous excitation within the probed sample volume.\\
\\
\textbf{Computational details.}
Density functional theory calculations are conducted within the plane-wave pseudopotential code {\tt Quantum ESPRESSO}~\cite{giannozzi_quantum_2009,giannozzi_advanced_2017},
using fully-relativistic norm-conserving pseudopotentials~\cite{hamann_optimized_2013} and the Perdew-Burke-Ernzerhof generalized-gradient approximation (PBE) for the exchange-correlation functional~\cite{perdew_generalized_1996}. The plane-wave kinetic-energy cutoff is set to 80~Ry, the Brillouin zone is sampled with a $10\times 6 \times 4$ Monkhorst-Pack grid, and the phonon dispersion is obtained from density functional perturbation theory on a $5\times 3\times 2$ q-point mesh~\cite{baroni_phonons_2001}. The crystal lattice vectors are chosen to reproduce the experimental ones from Ref.~\cite{mar_metal-metal_1992}, and the crystal structure is relaxed within that unit cell. The electron and phonon eigenvalues, as well as the electron-phonon coupling matrix $g_{nm}^{\rm \nu}({\bf k},{\bf q})$, are interpolated on a dense $40\times 24 \times 16$ grid via maximally-localized Wannier functions~\cite{marzari_maximally_2012} within the {\tt EPW} code~\cite{lee_electronphonon_2023}, which uses {\tt Wannier90} as a library~\cite{pizzi_wannier90_2020}.\\
\\
\textbf{\textit{Ab initio} description of coherent phonon excitation.}
The theoretical framework for the simulation of the coherent lattice dynamics is based on the solution of the coherent-phonon equation of motion~\cite{emeis_coherent_2025}:
\begin{align}
    \partial_t^2 Q_{\nu} + \gamma_\nu \partial_t Q_{\nu}
    + \omega^2_{ \nu} Q_{ \nu}
     = D_{\nu}(t) \quad .
    \label{eq:osci}
\end{align}
Here,  $\partial_t = \partial/\partial_t$, and $Q_{ \nu}$ is the adimensional displacement amplitude of the coherent phonon with frequency $\omega_{\nu}$. Also, $\gamma_\nu$ is the coherent-phonon damping rate mediated by phonon-phonon interactions~\cite{pan_origin_2025}. The room-temperature coherent phonon damping rate is estimated from the third-order force constant, which is obtained from finite differences using the {\tt third-order.py} utility of the {\tt shengBTE} code~\cite{li_shengbte_2014}.
The term on the right-hand side of Eq.~\ref{eq:osci} is the coherent-phonon driving force due to electron-phonon interaction:
\begin{align}\label{eq:D1}
D_{\nu}(t) = - \frac{2\omega_{\nu}}{\hbar} \sum_{n \mathbf{k}} g_{nn}^\nu (\mathbf{k},0)  \Delta f_{n\mathbf{k}}(t) \quad,
\end{align}
where $\Delta f_{n{\bf k}} = f_{n{\bf k}} (t)  - f_{n{\bf k}}^{(0)}$ is the change of the occupation function with $f_{n{\bf k}}^{(0)}$ as the (equilibrium) Fermi-Dirac distribution before excitation. The time-dependent electron distribution function is obtained within the time-dependent Boltzmann equations (TDBEs)~\cite{caruso_ultrafast_2022,pan_strain-induced_2024} as implemented in the {\tt EPW} code. The TDBEs are solved using an excited-state ansatz for the electrons described by a high-temperature Fermi-Dirac distribution $f_{n{\bf k}}(t=0) = [{\rm exp}((\varepsilon_{n{\bf k}}-\mu)/k_{\rm B}T_{\rm exc}) +1]^{-1}$, where the excitation temperature  $T_{\rm exc}$ is set to match the energy transferred to the system by an incident pump pulse with fluence $F$, as discussed in Ref.~\cite{emeis_coherent_2025}. The TDBE are solved through second-order Runge-Kutta time stepping with a time interval of 1~fs, within an energy window of 4~eV around the Fermi energy. 
The coherent-phonon amplitude obtained from Eqs.~\ref{eq:osci} and \ref{eq:D1} is illustrated in Fig.~S3a (single-pulse excitation). Based on this framework, the maximum  coherent-phonon amplitude  $Q_{\nu}^{\rm max}(F)$ is estimated by considering the static limit of Eq.~\ref{eq:osci} ($\partial_t Q^{\nu} = 0$) leading to $Q_{\nu}^{\rm max}(F) = D_{\nu}/ \omega^2_{ \nu}$. 
This value constitutes an upper theoretical limit to the maximum lattice displacement that the crystal can undergo for a given electronic excitation $\Delta f_{n\mathbf{k}}$. To simulate the coherent lattice dynamics induced by two distinct laser pulses, the coherent phonon equation of motion is solved by considering the action of two driving forces:
\begin{align}
D^{\nu}_{\rm dp}(t;\Delta t_{\text{p}_1-\text{p}_2}) = D_{\nu}(t) + D_{\nu}(t - \Delta t_{\text{p}_1-\text{p}_2}) 
\theta(\Delta t_{\text{p}_1-\text{p}_2})\quad, \label{eq:D2}
\end{align}
where $\Delta t_{p_1-p_2}$ is the time delay between the two pump pulses. 
The coherent-phonon amplitudes obtained from the time propagation of Eqs.\ref{eq:osci}-\ref{eq:D2} is illustrated in Figs.~S3~b-d for different values of the time delay $\Delta t_{\text{p}_1-\text{p}_2}$ between the two pulses. 
The frequency renormalization proportional to the coherent-phonon amplitude observed in experiments can be phenomenologically accounted for by substituting the bare phonon frequency $\omega_\nu$ with the dressed frequency $\Omega_\nu  = \omega_\nu -\beta |Q_\nu|$ in Eq.~ref{eq:osci}, where $\beta$ is a positive fitting parameter.\\
\\
\textbf{Symmetry relationships and calculation of the free-energy landscape.}
In the following, we again clarify the symmetry relationships between the different structural phases of WTe$_2$ and derive the Landau free-energy landscape depicted in Fig.\,\ref{fig:1}a. As elucidated in the main text, the polar Td phase belongs to the non-centrosymmetric space group $Pmn2_1$, while the non-polar 1T' phase crystallizes in the centrosymmetric $P2_1/m$ structure.  
Both can be derived from a higher-symmetry reference structure belonging to the $Pnma$ space group, which serves as their common supergroup.  
The relevant lattice instabilities of this $Pnma$ phase correspond to two irreducible representations: the polar interlayer sliding mode $\Gamma_2^-$ and the non-polar intralayer shear mode $\Gamma_4^+$.  
The $\Gamma_2^-$ distortion breaks inversion symmetry and lowers the symmetry to $Pmn2_1$, giving rise to the ferroelectric Td phase, while the $\Gamma_4^+$ distortion leads to monoclinic symmetry ($P2_1/m$), corresponding to the ferroelastic 1T' phase. We note that the interlayer sliding mode is often referred to in the literature as a `shear mode'. However, as evident from its irreducible representation ($\Gamma_2^-$), this mode does not break the symmetry of the elastic tensor and, therefore, is not a shear mode in the strict sense. In contrast, it is the intralayer shear mode ($\Gamma_4^+$) that is responsible for breaking an elastic symmetry.\\
\\
To describe the energetics associated with these structural distortions, we construct a Landau-type free-energy expansion of the $Pnma$ reference phase in terms of the mode amplitudes $Q_{\Gamma_2^-}$ and $Q_{\Gamma_4^+}$.  
Up to sixth order, the expansion reads
\begin{equation}
\begin{aligned}
F =\;& F_{Pnma}
+ a_{\Gamma_2^-}\, Q_{\Gamma_2^-}^2
+ b_{\Gamma_2^-}\, Q_{\Gamma_2^-}^4
+ c_{\Gamma_2^-}\, Q_{\Gamma_2^-}^6 \\
&+ a_{\Gamma_4^+}\, Q_{\Gamma_4^+}^2
+ b_{\Gamma_4^+}\, Q_{\Gamma_4^+}^4
+ c_{\Gamma_4^+}\, Q_{\Gamma_4^+}^6 \\
&+ \gamma\, Q_{\Gamma_2^-}^2 Q_{\Gamma_4^+}^2
+ \eta_{1}\, Q_{\Gamma_2^-}^4 Q_{\Gamma_4^+}^2
+ \eta_{2}\, Q_{\Gamma_2^-}^2 Q_{\Gamma_4^+}^4 \, .
\end{aligned}
\label{eq:Landau_expansion}
\end{equation}
\noindent Here, $Q_{\Gamma_2^-}$ is the amplitude of the polar interlayer sliding mode, which directly controls the out-of-plane ferroelectric polarization, while $Q_{\Gamma_4^+}$ represents the amplitude of the intralayer shear mode, associated with monoclinic distortion.  
The coupling coefficients $\gamma$, $\eta_1$, and $\eta_2$ describe nonlinear phonon–phonon interactions between the two order parameters and are essential for capturing the transition pathways between the $Pnma$ phase on the one hand and the Td or 1T' phases on the other hand. The fitted parameters entering Eq.\,\ref{eq:Landau_expansion} are summarized in Table\,\ref{tab:free_energy_parameters}. We compute the free energy from first principles using the VASP code~\cite{kresse_ab_1993} with the DFT+D3 van der Waals functional with Becke–Johnson damping~\cite{grimme_effect_2011}. The free-energy landscape was obtained by linearly interpolating the atomic positions between the relaxed $Pnma$, Td, and 1T' phases on a $11\times11$ grid, allowing the lattice parameters to relax at each interpolation step.
\begin{table}[h]
\centering
\begin{tabular}{l r}
\hline\hline
Parameter & Value [eV] \\
\hline
$a_{\Gamma_2^-}$     & $-2.00 \times 10^{-2}$ \\
$a_{\Gamma_4^+}$     & $-7.13 \times 10^{-3}$ \\
$b_{\Gamma_2^-}$     & $1.90 \times 10^{-2}$ \\
$\gamma$             & $5.90 \times 10^{-2}$ \\
$b_{\Gamma_4^+}$     & $3.72 \times 10^{-3}$ \\
$c_{\Gamma_2^-}$     & $1.41 \times 10^{-3}$ \\
$\eta_1$             & $-6.14 \times 10^{-3}$ \\
$\eta_2$             & $-1.53 \times 10^{-3}$ \\
$c_{\Gamma_4^+}$     & $1.73 \times 10^{-4}$ \\
\hline\hline
\end{tabular}
\caption{Fitted coefficients of the Landau free-energy expansion 
(up to sixth order) in terms of the order parameters 
$Q_{\Gamma_2^-}$ and $Q_{\Gamma_4^+}$. Energies are given in eV.}
\label{tab:free_energy_parameters}
\end{table}
\noindent The effective phonon potential discussed in the main text (see Fig.\,\ref{fig:1}a) is obtained by expanding the total free energy around its minimum at $(Q_{\Gamma_2^-}, Q_{\Gamma_4^+}) = (Q_{\Gamma_2^-}^{(0)}, 0)$, corresponding to the equilibrium configuration of the polar Td phase. This procedure allows us to isolate the anharmonic contributions governing the coupling between the ferroelectric and ferroelastic modes, and to map out the full free energy landscape connecting the Td, $Pnma$, and 1T' structures. In this configuration, any modulation of of the polar interlayer sliding mode $Q_{\Gamma_2^-}$ around its equilibrium value, that is \ $\delta Q_{\Gamma_2^-} = Q_{\Gamma_2^-} - Q_{\Gamma_2^-}^{(0)}$, transforms according to the totally symmetric $A_1$ irreducible representation of the $Pmn2_1$ space group. This means that near the Td minimum, the effective restoring potential for such fluctuations has $A_1$ symmetry, corresponding to the interlayer sliding mode that directly modulates the ferroelectric polarization $P_z$.

\section*{Declarations}

\begin{itemize}

\item \textbf{Acknowledgements.} This work was funded by an ETH Postdoctoral Fellowship, a Swiss National Science Foundation (SNSF) Postdoctoral Fellowship (TMPFP2 217303), and an ETH Seed Grant. The authors would like to thank M. Bauer, A. Lindenberg, S.L. Johnson, D. Soranzio, M. Savoini, and K.S. Burch for helpful discussions as well as J. Hecht, S. Reitz, and P. Moll for technical support.

\item \textbf{Author contributions}. The project was conceived by J.G.H. Experiments and data analysis were conducted by J.G.H., with contributions from A.C., N.W., and T.L. Q.M. carried out the symmetry analysis and related \textit{ab initio} calculations. C.E. and F.C. performed the \textit{ab initio} calculations on the energy-dependent electron-phonon coupling. The manuscript was written by J.G.H. M.F. provided overall guidance and support. All authors discussed the results and commented on the manuscript.

\item \textbf{Competing interests.} The authors declare no competing interests.

\item \textbf{Materials \& Correspondence}. Correspondence and requests for materials should be addressed to J.G.H.

\item \textbf{Data availability}. The data that support the findings of this study are available from the corresponding author upon reasonable request. 

\item \textbf{Code availability}. The code supporting the findings of this study is available from the corresponding author upon reasonable request.

\end{itemize}

\newpage

\newpage

\newpage

\newpage

\newpage

\begin{figure}[ht!]
    \centering
    \makebox[\textwidth][c]{
    \includegraphics[width=1.4\linewidth]{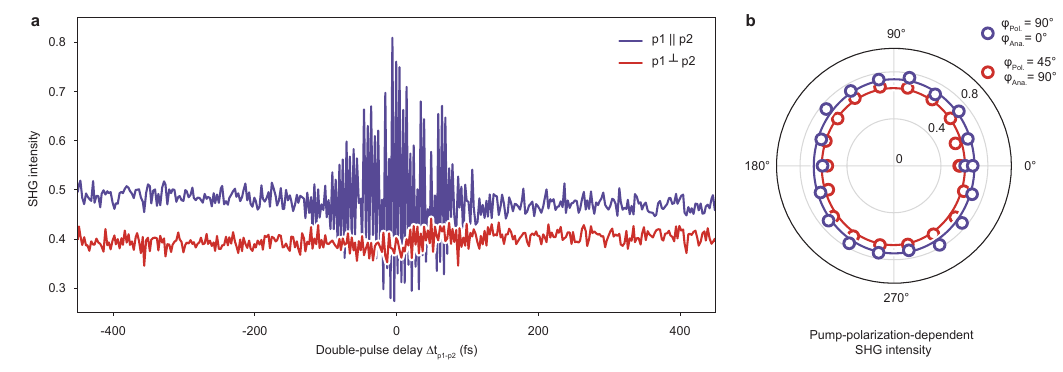}}
    \caption*{\textbf{Fig. S1 $\mid$  Polarization-dependent SHG signal in double-pulse and single-pulse experiments.} \textbf{a}, Comparison of SHG signals from  WTe$_2$ as a function of the delay between two pump pulses for parallel (violet) and cross-polarized (red) configurations (combined absorbed fluence $F=2.1\,\text{mJ}\,\text{cm}^{-2}$). The probe pulse interrogates the sample at a fixed delay of $\Delta t_{\text{p-pr}}= 100\,\text{ps}$, where coherent phonon oscillations are negligible. For parallel-polarized pump pulses, interference at the sample surface leads to a delay-dependent modulation of energy absorption, resulting in a corresponding variation of the SHG signal. In contrast, interference effects are strongly suppressed for cross-polarized pump pulses. Consequently, all double-pulse excitation experiments were performed using cross-polarized pump pulses. \textbf{b}, SHG intensity measured at $\Delta t_{\text{p-pr}}= 100\,\text{ps}$ for two different polarizer–analyzer configurations, shown as a function of the polarization angle of a single pump pulse (absorbed fluence $F=1.6\,\text{mJ}\,\text{cm}^{-2}$; $0^{\circ}: P\parallel a$-axis; $90^{\circ}: P\parallel b$-axis of the crystal). Only a weak anisotropy is observed, indicating that in double-pulse experiments both pump pulses, despite their differing polarizations, contribute similarly to the electronic excitation of the material.}
    \label{fig:S1}
\end{figure}

\newpage 

\begin{figure}[ht!]
    \centering
    \makebox[\textwidth][c]{
    \includegraphics[width=1.4\linewidth]{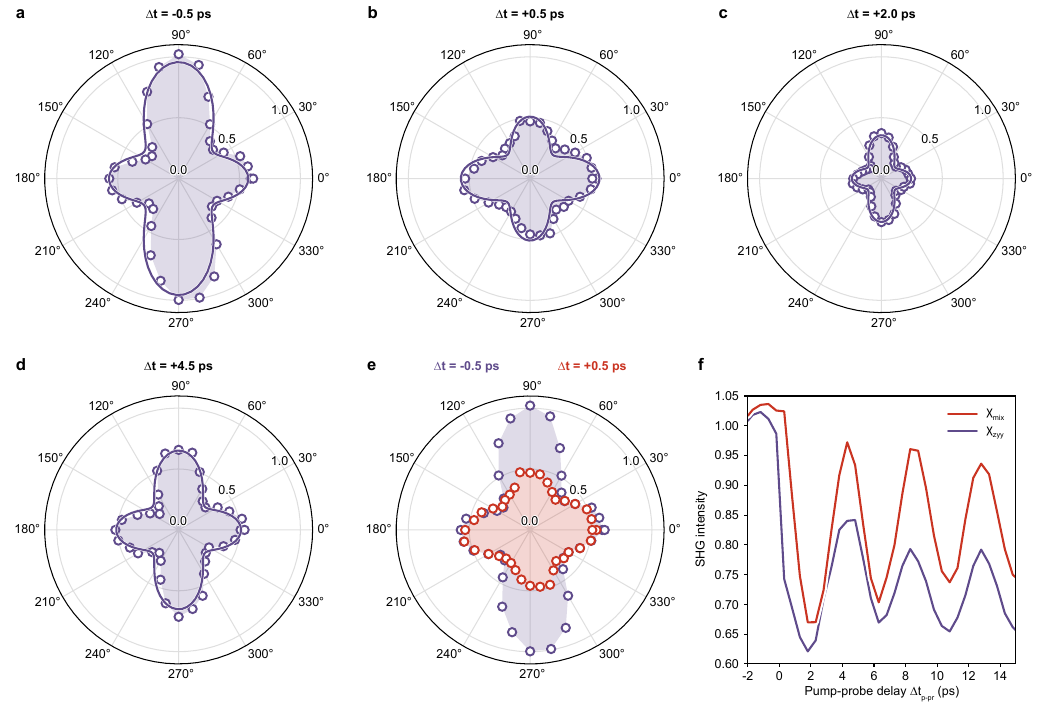}}
    \caption*{\textbf{Fig. S2 $\mid$  Time-dependent polarimetry measurements.} \textbf{a–d}, SHG intensity at selected pump–probe delays $\Delta t_{\text{p-pr}}$ as a function of the polarization angle of the fundamental light. The analyzer was fixed in the horizontal direction. Violet circles, experimental data; violet lines, fits based on the material’s symmetry (see Methods for details). \textbf{e}, Comparison of polarization-resolved SHG intensity just before and immediately after time zero. A pronounced change is observed in the $\chi_{zyy}$ component (at $90^\circ$), while the $\chi_{\text{mix}}$ component (at $0^\circ$) remains largely unaffected. \textbf{f}, Delay-dependent evolution of the SHG tensor components $\chi_{zyy}$ (violet) and $\chi_{\mathrm{mix}}$ (red), extracted from fits to the time-resolved polarimetry data. Upon photoexcitation, $\chi_{zyy}$ exhibits a rapid drop to a quasi-static level with superimposed oscillations at the sliding-mode frequency. In contrast, $\chi_{\mathrm{mix}}$ shows neither a fast initial suppression nor a comparable long-lived change.}
    \label{fig:S2}
\end{figure}

\newpage

\begin{figure}[ht!]
    \centering
    \makebox[\textwidth][c]{
    \includegraphics[width=0.6\linewidth]{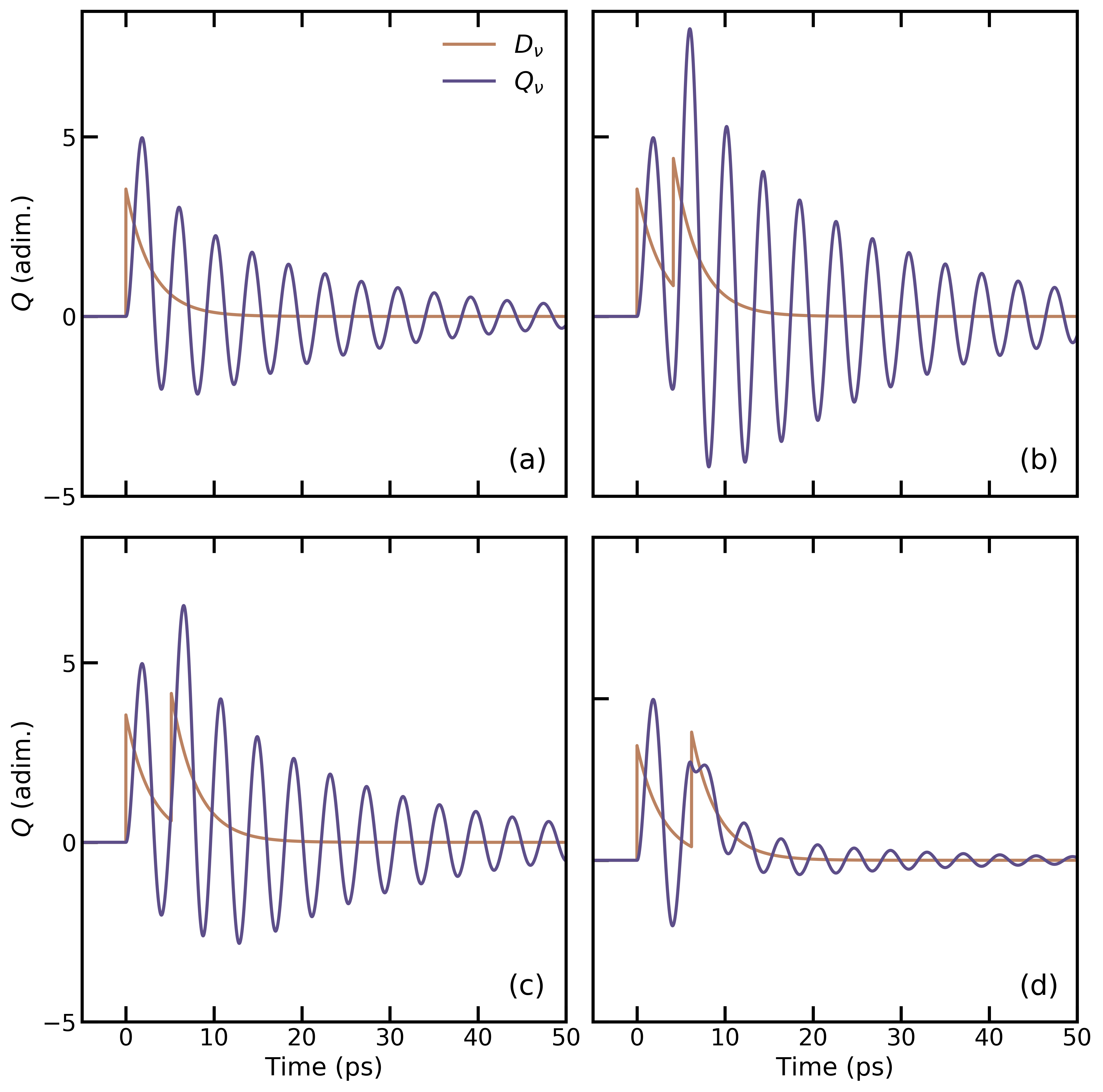}}
    \caption*{\textbf{ Fig. S3 $\mid$
    Time-resolved simulations of the coherent phonon excitation.}
    \textbf{a}, Time-dependent coherent phonon amplitude $Q_\nu$ for a single driving pulse, obtained from the solution of Eqs.~\ref{eq:osci}-\ref{eq:D1}, considering a single driving force marked in brown.
    \textbf{b}-\textbf{d}, Coherent phonon amplitude and driving force in the presence of two-pulse excitation as described via Eqs.~\ref{eq:osci}-\ref{eq:D2} for different time delays $\Delta t_{\text{p}_1-\text{p}_2}$ between the two pump pulses.
    Panel (b) illustrates the case in which the time delay coincides with one full phonon period, resulting in resonant excitation. In panels (c) and (d), the time delay is set to 1.25 and 1.5 phonon periods. In these cases, the second pulse results in the destructive interference of the coherent phonon, as revealed by the lower coherent phonon amplitude compared to the resonant case.
    }
    \label{fig:S3}
\end{figure}

\newpage

\begin{figure}[ht!]
    \centering
    \makebox[\textwidth][c]{
    \includegraphics[width=1.4\linewidth]{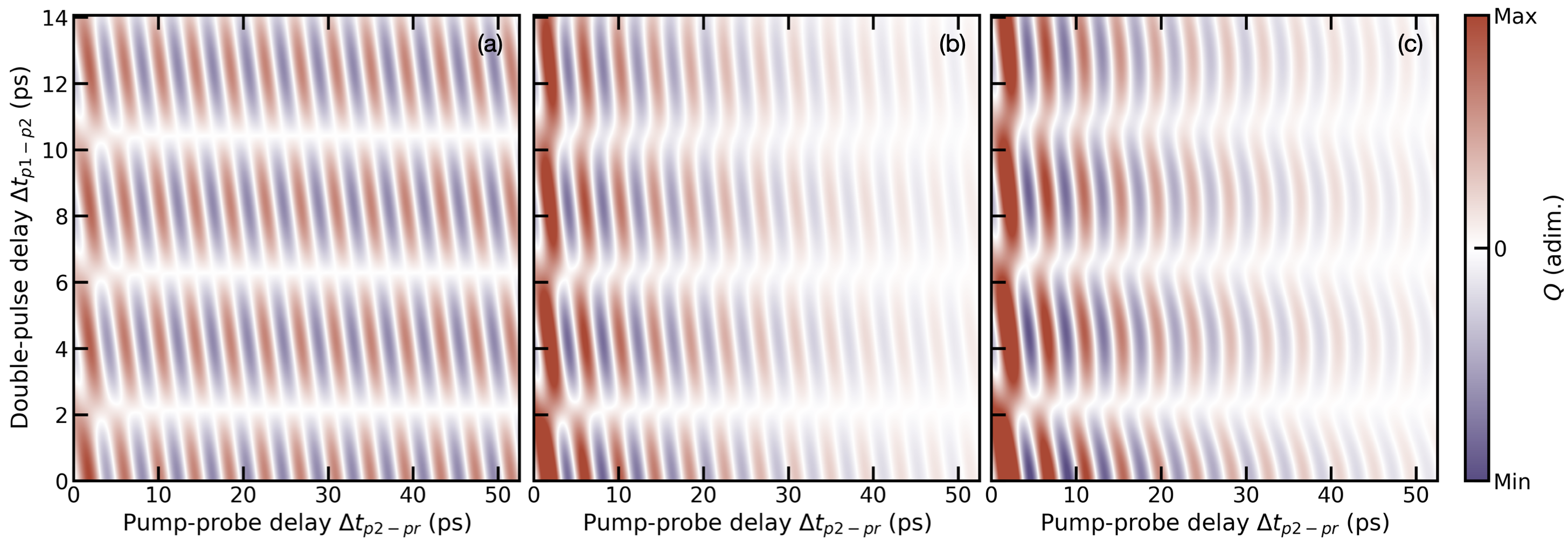}}
    \caption*{\textbf{ Fig. S4 $\mid$
    Double-pulse excitation of coherent phonons.} \textbf{a}-\textbf{c}, Coherent phonon amplitude as a function of time delay between two pump pulses  $\Delta t_{\text{p}_1-\text{p}_2}$ and of the pump-probe delay $\Delta t_{\text{p}_2-\text{p}_r}$. \textbf{a}, Coherent phonon amplitude obtained from Eq.~\ref{eq:osci} in absence of damping ($\gamma_\nu = 0$). \textbf{b}, Coherent phonon amplitude obtained from Eq.~\ref{eq:osci} in presence of damping. \textbf{c}, Coherent phonon amplitude obtained by considering the frequency renormalization according to $\Omega_\nu  = \omega_\nu -\beta |Q_\nu|$ in Eq.~\ref{eq:osci}.}
    \label{fig:S4}
\end{figure}

\newpage

\begin{figure}[ht!]
    \centering
    \makebox[\textwidth][c]{
    \includegraphics[width=1.4\linewidth]{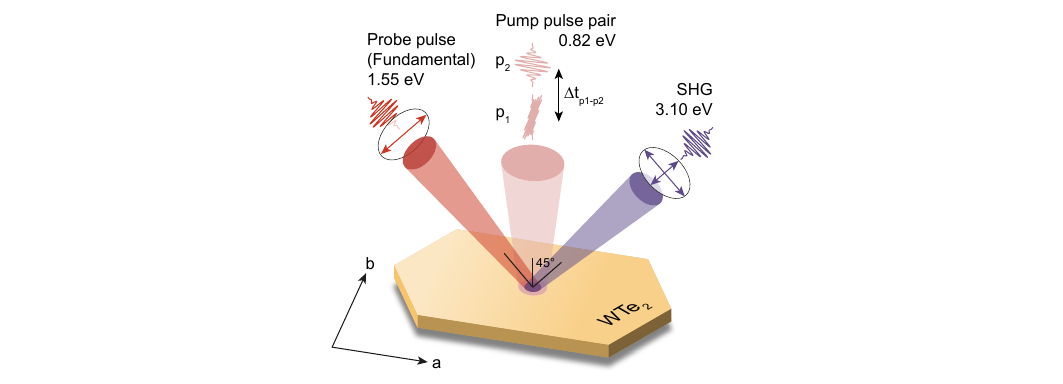}}
    \caption*{\textbf{ Fig. S5 $\mid$
    Experimental geometry.} Sketch of the experimental geometry for time-resolved SHG measurements with single- and double-pulse excitation. The coordinate system refers to the \textit{a}-axis and \textit{b}-axis of the WTe$_2$ crystal.}
    \label{fig:S5}
\end{figure}

\newpage

\begin{figure}[ht!]
    \centering
    \makebox[\textwidth][c]{
    \includegraphics[width=1.4\linewidth]{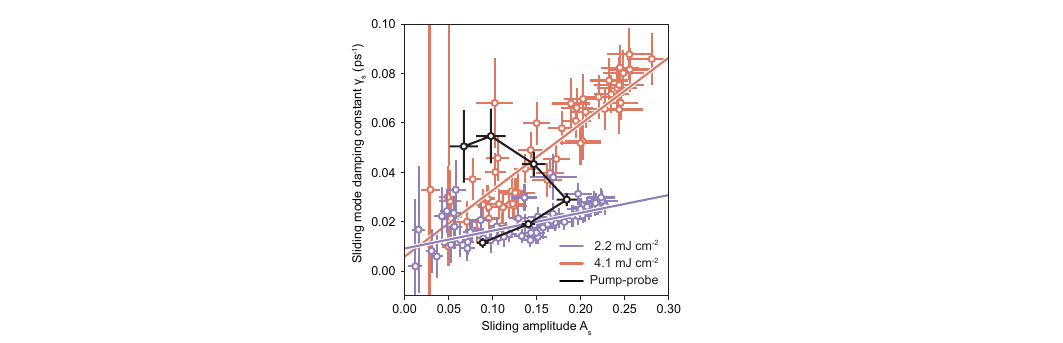}}
    \caption*{\textbf{ Fig. S6 $\mid$
    Comparison of amplitude-dependent damping in pump-probe and pump-pump-probe experiments.} \textbf{b}, Sliding-mode damping constant $\gamma_{\text{s}}$ as a function of the sliding-mode amplitude $A_{\text{s}}$ for two different combined fluences in double-pulse experiments (see also Fig.\,4b). Solid red and violet lines, linear models fitted to the data to determine  $(\Delta\gamma/\Delta A)_{\text{s}}$; errorbars, $1\sigma$-confidence intervals of the fits used to extract $A_{\text{s}}$ and $\gamma_{\text{s}}$. Black line and markers, values obtained in pump-probe experiments (see fluence-dependent sliding mode amplitude and damping in Fig. 1e). Notably, the linear relationship between the sliding-mode amplitude and its damping is evident only in the double-pulse data.}
    \label{fig:S6}
\end{figure}

%%%%%%%%%%%%%%%%%%%%%%%%%%%%%%%%


\newpage

\bibliography{WTe2_v3}% common bib file

\end{document}